\documentstyle[12pt,epsfig]{article}
\def\cite#1{#1}
\newcommand{\ct}[1]{[\cite{#1}]}
\def\thebibliography#1{\section*{References}\list
 {[\arabic{enumi}]}{\settowidth\labelwidth{[#1]}\leftmargin\labelwidth
 \advance\leftmargin\labelsep
 \usecounter{enumi}}
 \def\newblock{\hskip .11em plus .33em minus -.07em}
 \sloppy
 \sfcode`\.=1000\relax}

\begin{document}

\begin{center}
JLAB PROTON POLARIZATION DATA IN ASPECT TO GLOBAL UNITARY AND
ANALYTIC MODEL OF NUCLEON ELECTROMAGNETIC
STRUCTURE\footnote{Contribution presented at NAPP'03 Int.
Conference, May 26-31, 2003, Dubrovnik (Croatia)}
\end{center}

\begin{center}
S.~DUBNI\v{C}KA$^1$ and A.-Z.~DUBNI\v{C}KOV\'A$^2$
\end{center}

\begin{center} {
$^{1}$ \it Inst. of Physics, Slovak Acad. of Sci., Bratislava,
Slovak Republic\\
$^{2}$ \it Comenius Univ., Dept. of Theor.
Physics, Bratislava,
Slovak Republic }\\
\end{center}

\begin{abstract}
It is demonstrated that new JLAB proton polarization data, which
are in a rather strong disagreement with the proton electric form
factor data in the space-like region obtained by Rosenbluth
technique, are consistent with all known form factor properties,
including also the QCD asymptotics, but they require an existence
of a zero (i.e. a diffraction minimum) around $t$=15 GeV$^2$ of
the proton electric form factor. The latter leads to a change of
our knowledge about the charge distribution inside of the proton.
\end{abstract}


\section{Introduction}

The proton is charged particle compound of the ($u$, $u$, $d$)
quarks and therefore it is non-point like. As a consequence one
doesn't know an explicit form of the proton matrix element of the
electromagnetic (EM) current
\begin{equation}
J_{\mu}^{EM}=2/3 \bar u \gamma_{\mu} u -1/3 \bar d\gamma_{mu} d-1/3 \bar s \gamma_{\mu} s \label{a1}
\end{equation}
and therefore the latter is parametrized
\begin{equation}
\langle p|J_{\mu}^{EM}|p\rangle=\bar u(p')\left \{ \gamma_{\mu} F_{1p}(t)+i \frac{\sigma_{\mu\nu}q_{\nu}}{2m_p^2}F_{2p}(t)\right \} u(p) \label{a2}
\end{equation}
trough Dirac $F_{1p}(t)$ and Pauli $F_{2p}(t)$ form factors
(FF's), where  $t=q^2=(p'-p)^2=-Q^2$ is a four momentum
transferred by the virtual photon. From a practical point of view
it is advantageous to introduce Sach's FF's
\begin{eqnarray}
G_{Ep}(t) &=& F_{1p}(t)+\frac{t}{4m^2_p}F _{2p}(t) \label{a3} \\
G_{Mp}(t) &=& F_{1p}(t)+F_{2p}(t) \nonumber
\end{eqnarray}
to be normalized to the proton charge $G_{Ep}(0)=1$ and the proton magnetic moment $G_{Mp}(0)=1+\mu_p$
($\mu_p=1.793 [\mu]$ is the proton anomalous magnetic moment) and commonly they are called proton electric
and proton magnetic FF's, respectively.

Prior to the year 2000 all data on $G_{Ep}(t)$ and $G_{Mp}(t)$ in the space-like ($t<0$) region were
obtained by measuring (mainly at SLAC) the differential cross-section of elastic electron scattering on proton
\begin{eqnarray}
&&\frac{d\sigma^{lab}(e^-p\to e^-p)}{d\Omega}=\frac{\alpha^2}{4E^2}\frac{\cos^2(\theta/2)}{\sin^4(\theta/2)}\frac{1}{1+(\frac{2E}{m_p})\sin^2(\theta/2)}\cdot \label{a4} \\
&\cdot& \left [
\frac{G_{Ep}^2(t)-\frac{t}{4m_p^2}G_{Mp}^2(t)}{1-\frac{t}{4m_p^2}}-2\frac{t}{4m_{\pi}^2}G_{Mp}^2(t)\tan^2(\theta/2)\right
] \nonumber
\end{eqnarray}
and utilizing the Rosenbluth technique in order to separate
$G_{Ep}(t)$ and $G_{Mp}(t)$.

In such a way obtained data up to almost $t$=-35 GeV$^2$ have with increased $Q^2=-t$ smooth fall and the ratio
$(1+\mu_p)$$G_{Ep}(Q^2)/G_{Mp}(Q^2)$ has constant behavior at least up to $Q^2$=6 GeV$^2$, where the electric and
magnetic proton FF's follow the dipole formula \ct{1,2}
\begin{equation}
G_{Ep}(Q^2)=G_{Mp}(Q^2)/(1+\mu_p)=D(Q^2)=1/(1+\frac{Q^2}{071})^2. \label{a5}
\end{equation}
\begin{figure}[thb]
\begin{center}
\psfig{figure=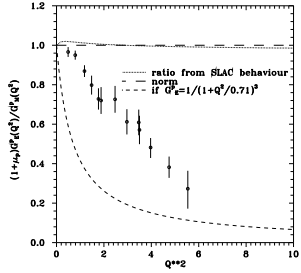,width=4cm}
\end{center}
\caption{Remarkable fall of $G_{Ep}(Q^2)$  with increased $Q^2$ in
comparison with $G_{Mp}(Q^2)$}
\end{figure}

More recently \ct{3,4} Jefferson Lab data on
$(1+\mu)G_{Ep}(Q^2)/G_{Mp}(Q^2)$, measuring simultaneously
transverse $P_t$ and longitudinal $P_l$ components of the recoil
proton's polarization in the electron scattering plane of the
polarization transfer process $\vec{e}p\to e\vec{p}$, have been
determined at the region $0.3 GeV^2\leq Q^2 \leq 5.6 GeV^2$ by the
relation
\begin{equation}
\frac{G_{Ep}}{G_{Mp}}=-\frac{P_t}{P_l}\frac{E+E'}{2m_p}\tan^2{(\theta/2)}, \label{a6}
\end{equation}
which reveal a remarkable fall of $G_{Ep}(Q^2)$ (see Fig.1) with
increased $Q^2$ in comparison with $G_{Mp}(Q^2)$ and so, this data
are in a rather strong disagreement with the data obtained by
Rosenbluth technique.

In this contribution we solve this puzzle by an investigation of a compatibility of new data with
all known FF properties in the framework of our global unitary and analytic model of the nucleon EM structure
\ct{5}.

\section{Global unitary and analytic model of nucleon  EM structure}

It is till now the most sophisticated model \ct{5} for four
independent analytic functions
\begin{eqnarray}
G_{Ep}(t)&=&[F_1^s(t)+F_1^v(t)]+\frac{t}{4m^2_p}[F_2^s(t)+F_2^v(t)]\nonumber \\
G_{Mp}(t)&=&[F_1^s+F_1^v(t)]+[F_2^s(t)+F_2^v(t)]\label{a7}
\\
G_{En}(t)&=&[F_1^s(t)-F_1^v(t)]+\frac{t}{4m^2_n}[F_2^s(t)-F_2^v(t)]\nonumber \\
G_{Mp}(t)&=&[F_1^s-F_1^v(t)]+[F_2^s(t)-F_2^v(t)],\nonumber
\end{eqnarray}
so-called electric and magnetic proton and neutron FF's, to be defined on four-sheeted Riemann surface in $t$-variable
with complex poles corresponding to unstable vector meson resonances placed only on unphysical
sheets. The model contains all known FF properties like experimental fact of a creation of vector-meson resonances
in electron-positron annihilation processes into hadrons, the assumed nucleon FF analytic properties,
unitary condition, normalization
\begin{equation}
F_1^s(0)=F_1^v(0)=1/2; \;\; F_2^s(0)=1/2(\mu_p+\mu_n); \;\;
F_2^v(0)=1/2(\mu_p-\mu_n), \label{a8}
\end{equation}
and the asymptotic behaviour as predicted by quark model of hadrons
\begin{equation}
F_1^{s,v}(t)_{|_{|t|\to \infty}}\sim  t^{-2}; \quad F_2^{s,v}(t)_{|_{|t|\to \infty}}\sim  t^{-3}.\label{a9}
\end{equation}
\begin{figure}[thb]
\begin{center}
\psfig{figure=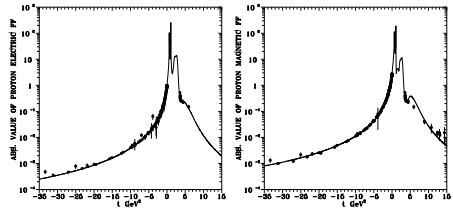,width=6cm}
\end{center}
\caption{Description of $G_{Ep}(t)$ and $G_{Mp}(t)$ data by
unitary and analytic model}
\end{figure}

\begin{figure}[thb]
\begin{center}
\psfig{figure=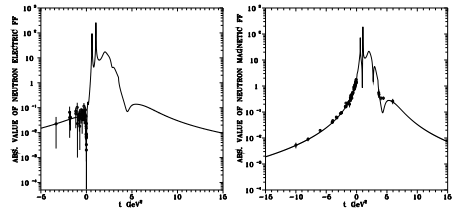,width=6cm}
\end{center}
\caption{Description of $G_{En}(t)$ and $G_{Mn}(t)$ data by
unitary and analytic model}
\end{figure}
In such a way it provides a very effective framework for a
consistent superposition of complex conjugate vector-meson pole
pair and continuum contributions with all other nucleon FF
properties and describes well (see Figs. 2,3) first time all
existing space-like (obtained by Rosenbluth technique) and
time-like nucleon FF data simultaneously, despite of the fact that
the obtained by Rosenbluth technique data on electric proton FF
$G_{Ep}(t)$ in $t<0$ region are significantly less precise than
the data on $G_{Mp}(t)$, since $G_{Mp}(t)$ is dominant in
differential cross-section (\ref{a4}), from the experimental
values of which both, $G_{Ep}(t)$ and $G_{Mp}(t)$, are extracted.

\section{Consequences following from JLAB proton polarization data}

From Fig.1 it is straightforward to see, that $G_{Ep}$
fall is steeper than the fall of dipole formula, but less intense than tripole one.

Knowing the ratio $G_{Ep}/G_{Mp}$ one can extract also data on the ratio $F_{2p}/F_{1p}$
by the expression
\begin{equation}
\frac{F_{2p}(Q^2)}{F_{1p}(Q^2)}=\left (1-\frac{G_{Ep}(Q^2)}{G_{Mp}(Q^2)}\right )/\left (\frac{G_{Ep}(Q^2)}{G_{Mp}(Q^2)}+\frac{Q^2}{4m^2_p}\right ).
\label{a10}
\end{equation}
\begin{figure}[thb]
\begin{center}
\psfig{figure=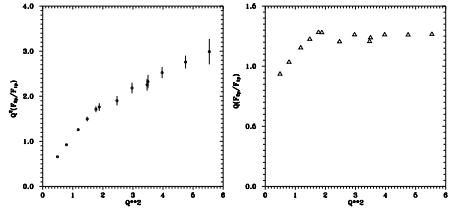,width=6cm}
\end{center}
\caption{Data on  $F_{2p}/F_{1p}$ multiplied by $Q^2$ and $Q$
respectively}
\end{figure}
Then $Q^2F_{2p}/F_{1p}$  indicates continuing increase (see Fig.4)
with $Q^2$. However, if we multiply the data on ratio
$F_{2p}/F_{1p}$ only by Q, then they acquire constant behaviour
with $Q^2$. So, the new JLab proton polarization data indicate
that the proton Pauli FF has (at least in the interval $1.8
GeV^2\leq Q^2\leq 5.6 GeV^2$) the behaviour
\begin{equation}
F_{2p}(Q^2)\sim Q^{-5} \label{a11}
\end{equation}
which is in contradiction with the PQCD predictions \ct{6},\ct{7}
\begin{equation}
F_{2p}(Q^2)_{|{|Q^2|\to \infty}}\sim Q^{-6}.
 \label{a12}
\end{equation}
If the latter is true, then from the definition of the proton
electric FF (\ref{a3}) it follows that $|G_{Ep}(Q^2)|$ must to
have a zero, i.e. diffraction minimum, at some higher $Q^2$.

\section{Solution of the puzzle}

On the base of our analysis above we are now in the position:
\begin{itemize}
\item[i)]
The data on $G_{Ep}(t)$ and $G_{Mp}(t)$ in $t<0$ region, obtained by
Rosenbluth technique from $d\sigma/d\Omega$,
are compatible with all other existing nucleon FF data and all known FF
properties, including also the QCD asymptotics.
\item[ii)]
The very precise JLab proton polarization data on
$(1+\mu_p)G_{Ep}(Q^2)/G_{Mp}(Q^2)$ contradict (at least in the
region $1.8\leq Q^2\leq 5.6 GeV^2$) predictions of PQCD.
\end{itemize}

Now, what data on $G_{Ep}(t)$ in $t<0$ region are true and what data are wrong?

In a solution of the latter problem we have proceeded as follows.
Since the magnetic proton FF $G_{Mp}(Q^2)$ is (owing to the factor
$\frac{t}{4m_p^2}$) dominant at the differential cross-section
(\ref{a4}) of $e^- p\to e^- p$ (at $Q^2$=$-t\approx 3GeV^2$ the
electric proton FF contributes only 5$\%$ to the cross-section
value and for higher momenta it is even less), the obtained data
on electric proton FF could be unreliable, even wrong. On the
other hand, a simultaneous measurement of the transverse and
longitudinal components of the recoil proton's polarization in the
electron scattering plane of the polarization transfer process
$\vec{e}^- p\to e^-\vec{p}$ is very effective and reliable method
of a determination of the ratio $G_{Ep}/G_{Mp}$. As a result we
believe in the obtained data presented in Fig.1 and their
violation of QCD asymptotics given by (\ref{a11}) we consider to
be a local effect.

This our conjecture we verify in the framework of the global
unitary and analytic model of nucleon EM structure which contains
the QCD asymptotics automatically. First, we exclude from the set
of all existing proton and neutron EM FF data the space-like data
on $G_{Ep}(t)$ obtained from $d\sigma/d\Omega$ by Rosenbluth
technique. We substitute them for JLab proton polarization data on
$(1+\mu_p)G_{Ep}(t)/G_{Mp}(t)$. Then by means of the 10 -resonance
unitary and analytic model of nucleon EM structure \ct{5} we carry
out a fitting procedure of all these data simultaneously which
provides in such a way test of:
\begin{itemize}
\item[i)]
consistency of the JLab proton polarization data with all other proton and neutron EM FF data
\item[ii)]
consistency of these data with the powerful tool of physics - the analyticity
\item[iii)]
their consistency with the asymptotic behaviour as predicted (up to logarithmic corrections)
by QCD.
\end{itemize}
\begin{figure}[thb]
\begin{center}
\psfig{figure=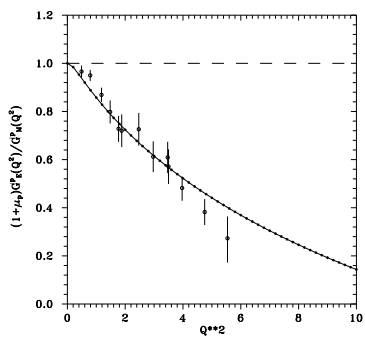,width=5cm}
\end{center}
\caption{Description of the  JLab proton polarization data by
unitary and analytic model}
\end{figure}

\begin{figure}[hb]
\begin{center}
\psfig{figure=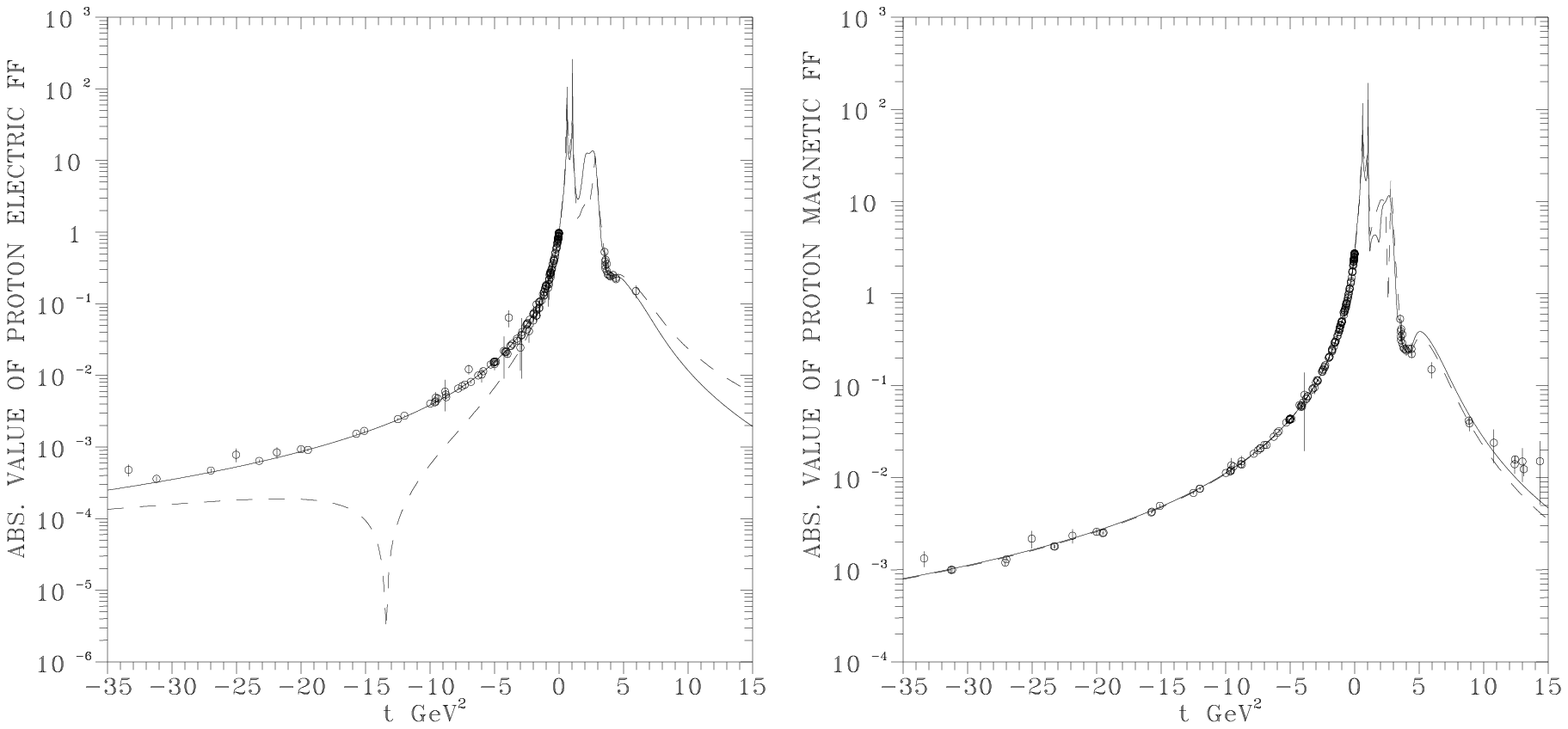,width=7cm}
\end{center}
\caption{Results of the JLab proton $t<0$ data analysis (dashed
lines) for $G_{Ep}(t)$ and $G_{Mp}(t)$}
\end{figure}

\begin{figure}[thb]
\begin{center}
\psfig{figure=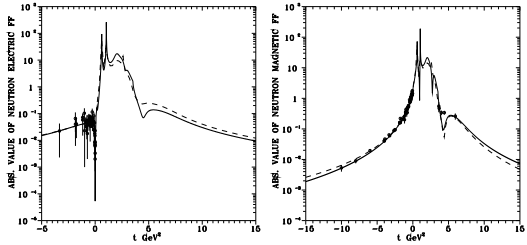,width=7cm}
\end{center}
\caption{Results of the JLab proton $t<0$ data analysis (dashed
lines) for $G_{En}(t)$ and $G_{Mn}(t)$}
\end{figure}
 We have obtained very surprising results. A perfect description
of the JLab proton polarization data (see Fig.5) was achieved. The
fitted parameters of the unitary and analytic model of nucleon EM
structure are almost unchanged in comparison with those given in
\ct{5}. A description of the $G_{Ep}(t)$ time-like data, of all
$G_{Mp}(t)$ data and space-like and time-like neutron EM FF data
is (see dashed lines in Figs. 6,7) changed very little.

So, we come to the conclusion, that the JLab proton polarization
data are consistent with all other existing nucleon EM FF data
besides space-like $G_{Ep}(t)$ data obtained from $d\sigma(e^-
p\to e^- p)/d\Omega$ by Rosenbluth technique. They are consistent
with analyticity and they don't contradict the QCD asymptotics,
however they strongly require an existence of the zero in
$|G_{Ep}(t)|$ (i.e. a diffraction minimum well known in nuclear EM
FF's) around $t=-15 GeV^2$.

\section{Conclusions and discussion}

Recent JLab proton polarization data on $(1+\mu_p)G_{Ep}(Q^2)/G_{Mp}(Q^2)$,
 obtained by a simultaneous measurement of the transverse and longitudinal
 components of the recoil proton's polarization in the electron scattering
plane of the polarization transfer process $\vec{e} p\to e\vec{p}$
at the region $0.3 GeV^2\leq Q^2\leq 5.6 GeV^2$, revealed a
remarkable fall of $G_{Ep}(Q^2)$ with increased
$Q^2$, which
contradicts previous $G_{Ep}(t)$ data determined from
$d\sigma(e^-p\to e^-p)/d\Omega$ by Rosenbluth technique and so,
also the QCD asymptotics.

Discarding all $G_{Ep}(t)$ data obtained by Rosenbluth technique
in space-like region we have analyzed new JLab proton polarized
space-like data together with all other nucleon EM FF data in the
framework of the unitary and analytic model of the nucleon EM
structure from Ref. \ct{5}.

On the basis of the obtained results we came to the conclusions,
that the JLab proton polarization data are consistent with all
nucleon EM FF data, besides space-like $G_{Ep}(t)$ data obtained
by Rosenbluth technique, and also with QCD asymptotics, however,
they require an existence of a zero (i.e. diffraction minimum) of
$|G_{Ep}(t)|$ around $t=-15 GeV^2$. So there is a challenge to
experimental groups to confirm our conclusion by measuring
$(1+\mu_p)G_{Ep}(Q^2)/G_{Mp}(Q^2)$ at higher momenta.

This work was in part supported by Slovak Grant Agency for
Sciences, Grant 2/1111/23.

\end{document}